\documentclass[12pt,aps,prb,preprint]{revtex4}   
\usepackage{amsmath}    
\usepackage{graphicx}   
\begin{document}
\title{Comment on ``Exact Expression for Radiation of an Accelerated Charge in Classical Electrodynamics''}
\author{Ashok K. Singal}
\affiliation{Astronomy and Astrophysics Division, Physical Research Laboratory,
Navrangpura, Ahmedabad - 380 009, India }
\email{asingal@prl.res.in}
\begin{abstract}
It is shown that a newly derived ``exact expression'' 
for radiation of an accelerated charge in the recent literature is simply incorrect, 
having arisen because of a wrong relativistic transformation of the distance parameter. 
The ensuing claim that the newly derived expression alone that  
satisfies the energy conservation for the electromagnetic radiation, is based 
on a wrong reasoning where a proper distinction between the time during which the 
radiation is received and the time for emission (retarded time of 
the charge) was not maintained. 

\end{abstract}
\maketitle
\section{Introduction}
Huang and Lu  [\onlinecite{7}] claim to have derived new ``exact expression'' 
for radiation of an accelerated charge in classical electrodynamics. 
and in a later paper Huang [\onlinecite{8}] has claimed that it is their newly derived expression alone that  
satisfies the energy conservation for the electromagnetic radiation. On the other hand a proper 
transformation for all quantities does lead us to the standard expression for the electric field [\onlinecite{6}].
Here we explore the cause of the discrepancy to decide which of the two alernatives are the correct 
expressions for the electric and magnetic fields, leading to the correct expression for the 
angular distribution of radiation in classical electrodynamics.
\section{Discrepancy in expressions for the electric and magnetic fields}
The discrepancy in expressions for angular distribution of radiation in the standard approach [\onlinecite{1,2,3}] 
and in that by Huang and Lu [\onlinecite{7}] arises because in the latter they did not properly 
transform $r'$ and $r$, the distance between the field point and the retarded position of the charge as 
specified in the two frames. While deriving expressions for the electric and magnetic fields, ${\bf E}$ and ${\bf B}$, 
 Huang and Lu  [\onlinecite{7}] in their Eqs. (20-25) simply replaced $r'$ with $r$ which is not correct as these  
two quantities are actually related by $r' = r/\delta$, where $\delta = 1/\gamma (1-{{\bf \hat r}}\;.\;\mbox{\boldmath $\beta$})$
is the Doppler factor [\onlinecite{6}]. Thus their transformed electric and magnetic fields are wrong. 
Though they did employ the correct transformations for $\sin \theta$ and $\cos \theta$ 
(their Eqs. (15-16)), yet the related transformation between $r'$ and $r$ was strangely ignored.
It does not appear to be an inadvertent omission as they have later justified their formulation to be the 
only correct one [\onlinecite{7,8}].
Their derived expressions for the electric and magnetic fields (in SI units)    
\begin{equation}
{\bf E}=\frac {q}{4\pi\epsilon_0c} \,\left[
 \frac{\gamma{{\bf \hat r}}\times\{({{\bf \hat r}}-\mbox{\boldmath $\beta$})\times
\dot{\mbox{\boldmath $\beta$}}\}}{r\,(1-{{\bf \hat r}}\;.\;\mbox{\boldmath $\beta$})^{2}} \right]
\end{equation}
\begin{equation}
{\bf B}= \frac {q\mu_0}{4\pi} \,\left[
 \frac{\gamma {\bf \hat r}\times ({{\bf \hat r}}\times\{({{\bf \hat r}}-\mbox{\boldmath $\beta$})\times
\dot{\mbox{\boldmath $\beta$}}\})}{r\,(1-{{\bf \hat r}}\;.\;\mbox{\boldmath $\beta$})^{2}} \right]
\end{equation}
reduce to the standard expressions, if we replace $r$ with $r \gamma (1-{{\bf \hat r}}\;.\;\mbox{\boldmath $\beta$})$.  
It should be noted that throughout all quantities in the square brackets are to be evaluated at the retarded time.

Though Huang and Lu  [\onlinecite{7}] did not derive expressions for the velocity fields, but if we do follow their approach of 
using $r'=r$ for the velocity fields as well, then we end up with an electric field,
\begin{equation}
{\bf E}=\frac {q}{4\pi\epsilon_0} \,\left[
\frac{({\bf \hat r}-\mbox{\boldmath $\beta$})
}{r^2\,(1-{{\bf \hat r}}\;.\;\mbox{\boldmath $\beta$})}\right]
\end{equation}
with a $\delta^2$ term missing with respect to the standard expression for the velocity fields, 
which could be derived even otherwise from a direct Lorentz transformation of the rest-frame Coulomb field, [\onlinecite{1,2}] 
again showing that their approach of putting $r'=r$ is incorrect.

On the other hand a proper transformation for all quantities does lead us [\onlinecite{6}] to the standard expression 
[\onlinecite{1,2,3}] for the fields, 
\begin{equation}
{\bf E}=\frac {q}{4\pi\epsilon_0} \,\left[
\frac{({\bf \hat r}-\mbox{\boldmath $\beta$})
}{r^2\,\gamma^2(1-{{\bf \hat r}}\;.\;\mbox{\boldmath $\beta$})^{3}}\right]
+\frac {q}{4\pi\epsilon_0c} \, \,\left[
 \frac{{{\bf \hat r}}\times\{({{\bf \hat r}}-\mbox{\boldmath $\beta$})\times
\dot{\mbox{\boldmath $\beta$}}\}}{r\,(1-{{\bf \hat r}}\;.\;\mbox{\boldmath $\beta$})^{3}} \right]
\end{equation}
both for velocity fields as well as radiation fields, with magnetic field 
${\bf B}= {\bf \hat r}\times {\bf E}/{c}$.
\section{angular distribution of the received versus emitted radiated power}
From their derived electric and magnetic field expressions, Huang and Lu  [\onlinecite{7,8}] then got the Poynting 
vector as
\begin{equation}
{\bf S}=\frac{{\bf E}\times {\bf B}}{\mu_0}=\frac {q^2}{16\pi^2\epsilon_0c} \,\left[
 \frac{\gamma^2({{\bf \hat r}}\times\{({{\bf \hat r}}-\mbox{\boldmath $\beta$})\times
\dot{\mbox{\boldmath $\beta$}}\})^2}{r^2\,(1-{{\bf \hat r}}\;.\;\mbox{\boldmath $\beta$})^{4}} \right]\,,
\end{equation}
and after evaluating the radiation crossing a spherical surface surrounding the charge in a time interval $dt$, they 
equated it to the power emitted by the charge in the time interval $dt$. This however is fallacious.
\begin{figure}
\scalebox{0.8}{\includegraphics{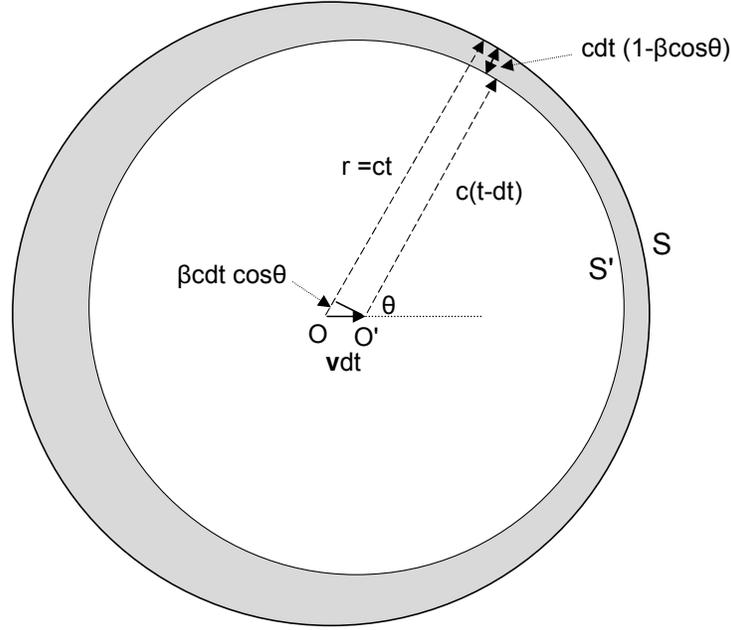}}
\caption{Radiation from a moving charge emitted during a time interval $dt$ lies in 
the region of radial width $cdt(1-\beta \cos \theta)$, 
enclosed within the spherical surfaces $S$ and $S'$, indicated by the darker shade.}
\end{figure}

For a charge moving with velocity ${\bf v}$, the spherical surface $S$ at $r=ct$ may be centered on $O$, but in time $dt$
the charge moves to $O'$, a distance ${\bf v}dt$ away from $O$, and no longer remains at the center of $S$. 
The energy radiated during the time interval $dt$ lies  
inside of the spherical surface $S$ (of radius $r=ct$ centered on $O$) and outside of $S'$ (of radius $c(t-dt)$ 
centered at $O'$). Thus the radiation emitted during $dt$ lies in a region enclosed within surface $S$ 
and $S'$ and is not distributed with spherical symmetry around $O$, basically as the emitting charge did not remain  
stationed at $O$ during this time interval $dt$.
As can be readily seen from Fig.~1, the radial width of the zone of radiation is not a constant $c dt$ but instead varies 
with $\theta$ as $cdt(1-\beta \cos \theta)$.
Therefore the radiation emitted by the charge during a time interval $dt$ does not cross the spherical surface $S$ 
at all points in a time interval $dt$  but instead, depending upon $\theta$, it takes 
$dt(1-\beta \cos \theta)=dt(1-{{\bf \hat r}}\;.\;\mbox{\boldmath $\beta$})$. 
And that leads us to the correct expression 
for the total power radiated (Li\'{e}nard's result) [\onlinecite{1,2,3}]
\begin{equation}
P=\frac {q^2}{6\pi\epsilon_0c} \,\left[\gamma^6\{\dot{\mbox{\boldmath $\beta$}}^2
 -(\mbox{\boldmath $\beta$}\times
\dot{\mbox{\boldmath $\beta$}})^2\} \right].
\end{equation}
Here it is not a Lorentz transformation of time that is involved, as we are considering only the observer's frame in which the charge 
has a velocity ${\bf v}$; it is merely a matter of simple geometry in that frame and which cannot be overlooked. 
Therefrore taking the geometry into account, 
the radiation formula derived by Huang and Lu  [\onlinecite{7}] in fact will not lead to a 
correct result for the total power radiated. It is only the standard expressions for the electric and magnetic fields and  
thereby derived radiation formula which yields the correct value for the total radiated power.
\section{Conclusion}
We have shown that the so-called exact expressions for the electric and magnetic fields of an accelerated 
charge derived recently in the literature are incorrect as these resulted from an incorrect relativistic transformation 
of the distance parameter. Also the derived formulation for the angular distribution of radiation does not lead to a 
correct result for the total power radiated when a proper distinction is made between the time during which the 
radiation is received and the time of emission (retarded time) of 
the charge. 

\end{document}